\documentclass[12pt,a4paper]{article}

\usepackage{ifthen} 
\newboolean{pdflatex}
\setboolean{pdflatex}{true} 

\newboolean{articletitles}
\setboolean{articletitles}{true} 

\newboolean{uprightparticles}
\setboolean{uprightparticles}{false} 

\newboolean{inbibliography}
\setboolean{inbibliography}{false} 

\usepackage{microtype}
\usepackage{lineno}  
\usepackage{xspace} 
\usepackage{caption} 

\usepackage{graphicx}  
\usepackage{color}
\usepackage{colortbl}
\graphicspath{{./figs/}} 

\usepackage{amsmath} 
\usepackage{amssymb}
\usepackage{amsfonts}
\usepackage{upgreek} 

\newcommand*\patchAmsMathEnvironmentForLineno[1]{%
\expandafter\let\csname old#1\expandafter\endcsname\csname #1\endcsname
\expandafter\let\csname oldend#1\expandafter\endcsname\csname
end#1\endcsname
 \renewenvironment{#1}%
   {\linenomath\csname old#1\endcsname}%
   {\csname oldend#1\endcsname\endlinenomath}%
}
\newcommand*\patchBothAmsMathEnvironmentsForLineno[1]{%
  \patchAmsMathEnvironmentForLineno{#1}%
  \patchAmsMathEnvironmentForLineno{#1*}%
}
\AtBeginDocument{%
\patchBothAmsMathEnvironmentsForLineno{equation}%
\patchBothAmsMathEnvironmentsForLineno{align}%
\patchBothAmsMathEnvironmentsForLineno{flalign}%
\patchBothAmsMathEnvironmentsForLineno{alignat}%
\patchBothAmsMathEnvironmentsForLineno{gather}%
\patchBothAmsMathEnvironmentsForLineno{multline}%
\patchBothAmsMathEnvironmentsForLineno{eqnarray}%
}

\usepackage[colorlinks=true,citecolor=blue]{hyperref}    
\usepackage[all]{hypcap} 

\def\MagUp {\mbox{\em Mag\kern -0.05em Up}\xspace}

\ifthenelse{\boolean{uprightparticles}}%
{

 \def\Ppi         {\ensuremath{\uppi}\xspace}

 \def\PDelta      {\ensuremath{\Delta}\xspace}                 
 \def\PXi      {\ensuremath{\Xi}\xspace}                 
 \def\PLambda      {\ensuremath{\Lambda}\xspace}                 
 \def\PSigma      {\ensuremath{\Sigma}\xspace}                 
 \def\POmega      {\ensuremath{\Omega}\xspace}                 
 \def\PUpsilon      {\ensuremath{\Upsilon}\xspace}                 
 
 \def\PB      {\ensuremath{\mathrm{B}}\xspace}                 
                  
 \def\PD      {\ensuremath{\mathrm{D}}\xspace}

 \def\PK      {\ensuremath{\mathrm{K}}\xspace}

 \def\Pi      {\ensuremath{\mathrm{i}}\xspace}

}
{

 \def\Ppi         {\ensuremath{\pi}\xspace}

 \mathchardef\PDelta="7101
 \mathchardef\PXi="7104
 \mathchardef\PLambda="7103
 \mathchardef\PSigma="7106
 \mathchardef\POmega="710A
 \mathchardef\PUpsilon="7107
                  
 \def\PB      {\ensuremath{B}\xspace}                 
                  
 \def\PD      {\ensuremath{D}\xspace}

 \def\PK      {\ensuremath{K}\xspace}

 \def\Pi      {\ensuremath{i}\xspace}

}

\makeatletter
\ifcase \@ptsize \relax
  \newcommand{\miniscule}{\@setfontsize\miniscule{4}{5}}
\or
  \newcommand{\miniscule}{\@setfontsize\miniscule{5}{6}}
\or
  \newcommand{\miniscule}{\@setfontsize\miniscule{5}{6}}
\fi
\makeatother

\DeclareRobustCommand{\optbar}[1]{\shortstack{{\miniscule (\rule[.5ex]{1.25em}{.18mm})}
  \\ [-.7ex] $#1$}}

\def\pion   {{\ensuremath{\Ppi}}\xspace}
\def\piz    {{\ensuremath{\pion^0}}\xspace}

\def\pip    {{\ensuremath{\pion^+}}\xspace}
\def\pim    {{\ensuremath{\pion^-}}\xspace}
\def\pipm   {{\ensuremath{\pion^\pm}}\xspace}

\def\kaon    {{\ensuremath{\PK}}\xspace}
  \def\Kbar    {{\kern 0.2em\overline{\kern -0.2em \PK}{}}\xspace}

\def\KorKbar    {\kern 0.18em\optbar{\kern -0.18em K}{}\xspace}

\def\Kp      {{\ensuremath{\kaon^+}}\xspace}
\def\Km      {{\ensuremath{\kaon^-}}\xspace}
\def\Kpm     {{\ensuremath{\kaon^\pm}}\xspace}
\def\Kmp     {{\ensuremath{\kaon^\mp}}\xspace}
\def\KS      {{\ensuremath{\kaon^0_{\rm\scriptscriptstyle S}}}\xspace}

\def\Kstarz  {{\ensuremath{\kaon^{*0}}}\xspace}

\def\Kstarp  {{\ensuremath{\kaon^{*+}}}\xspace}
\def\Kstarm  {{\ensuremath{\kaon^{*-}}}\xspace}

  \def\Dbar    {{\kern 0.2em\overline{\kern -0.2em \PD}{}}\xspace}
\def\D       {{\ensuremath{\PD}}\xspace}

\def\DorDbar    {\kern 0.18em\optbar{\kern -0.18em D}{}\xspace}
\def\DtwoorDtwobar {\kern -0.25em\optbar{\kern 0.25em D_2^*}{}\xspace}
\def\Dz      {{\ensuremath{\D^0}}\xspace}
\def\Dzb     {{\ensuremath{\Dbar{}^0}}\xspace}

\def\Dstar   {{\ensuremath{\D^*}}\xspace}

\def\B       {{\ensuremath{\PB}}\xspace}
\def\Bbar    {{\ensuremath{\kern 0.18em\overline{\kern -0.18em \PB}{}}}\xspace}

\def\BorBbar    {\kern 0.18em\optbar{\kern -0.18em B}{}\xspace}
\def\Bz      {{\ensuremath{\B^0}}\xspace}

\def\BzorBzbar  {\kern 0.18em\optbar{\kern -0.18em B}{}^0\xspace}
\def\Bu      {{\ensuremath{\B^+}}\xspace}

\def\Bp      {{\ensuremath{\Bu}}\xspace}

\def\Bpm     {{\ensuremath{\B^\pm}}\xspace}

  \def\Y#1S{\ensuremath{\PUpsilon{(#1S)}}\xspace}

\def\Lbar        {{\ensuremath{\kern 0.1em\overline{\kern -0.1em\PLambda}}}\xspace}
\def\LorLbar    {\kern 0.18em\optbar{\kern -0.18em \PLambda}{}\xspace}

\def\to                 {\ensuremath{\rightarrow}\xspace}

\def\CP                {{\ensuremath{C\!P}}\xspace}

\def\AT#1     {\ensuremath{A_{\mathrm{T}}^{#1}}\xspace}           

\def\C#1      {\ensuremath{\mathcal{C}_{#1}}\xspace}                       
\def\Cp#1     {\ensuremath{\mathcal{C}_{#1}^{'}}\xspace}                    
\def\Ceff#1   {\ensuremath{\mathcal{C}_{#1}^{\mathrm{(eff)}}}\xspace}        
\def\Cpeff#1  {\ensuremath{\mathcal{C}_{#1}^{'\mathrm{(eff)}}}\xspace}       
\def\Ope#1    {\ensuremath{\mathcal{O}_{#1}}\xspace}                       
\def\Opep#1   {\ensuremath{\mathcal{O}_{#1}^{'}}\xspace}                    

\newcommand{\tev}{\ifthenelse{\boolean{inbibliography}}{\ensuremath{~T\kern -0.05em eV}\xspace}{\ensuremath{\mathrm{\,Te\kern -0.1em V}}}\xspace}
\newcommand{\gev}{\ensuremath{\mathrm{\,Ge\kern -0.1em V}}\xspace}
\newcommand{\mev}{\ensuremath{\mathrm{\,Me\kern -0.1em V}}\xspace}
\newcommand{\kev}{\ensuremath{\mathrm{\,ke\kern -0.1em V}}\xspace}
\newcommand{\ev}{\ensuremath{\mathrm{\,e\kern -0.1em V}}\xspace}
\newcommand{\gevc}{\ensuremath{{\mathrm{\,Ge\kern -0.1em V\!/}c}}\xspace}
\newcommand{\mevc}{\ensuremath{{\mathrm{\,Me\kern -0.1em V\!/}c}}\xspace}
\newcommand{\gevcc}{\ensuremath{{\mathrm{\,Ge\kern -0.1em V\!/}c^2}}\xspace}
\newcommand{\gevgevcccc}{\ensuremath{{\mathrm{\,Ge\kern -0.1em V^2\!/}c^4}}\xspace}
\newcommand{\mevcc}{\ensuremath{{\mathrm{\,Me\kern -0.1em V\!/}c^2}}\xspace}

\def\gsim{{~\raise.15em\hbox{$>$}\kern-.85em
          \lower.35em\hbox{$\sim$}~}\xspace}
\def\lsim{{~\raise.15em\hbox{$<$}\kern-.85em
          \lower.35em\hbox{$\sim$}~}\xspace}

\def\degrees{\ensuremath{^{\circ}}\xspace}

\def\tell1  {TELL1\xspace}
\def\ukl1   {UKL1\xspace}

\newcommand{\eg}{\mbox{\itshape e.g.}\xspace}
\newcommand{\ie}{\mbox{\itshape i.e.}\xspace}

\usepackage{cite} 
\usepackage{mciteplus}

\usepackage{rotating} 
\usepackage{arydshln} 
\usepackage{relsize} 
\usepackage[hang,flushmargin]{footmisc} 

\makeatletter
\g@addto@macro\bfseries{\boldmath}
\makeatother

\begin{document}

\renewcommand{\thefootnote}{\fnsymbol{footnote}}
\setcounter{footnote}{1}

\begin{titlepage}
\pagenumbering{roman}

\begin{center}
\bf\huge
Origins of the method to determine the CKM angle $\gamma$ using $\Bpm \to D\Kpm$, $D \to \KS\pip\pim$ decays
\end{center}

\vspace*{1.5cm}

\begin{center}
%
  A.~Ceccucci$^c$, T.~Gershon$^g$, M.~Kenzie$^g$, Z.~Ligeti$^l$, Y.~Sakai$^s$, K.~Trabelsi$^t$
\bigskip\\
{\it\footnotesize 
$ ^c$ EP Department, CERN, Switzerland\\
$ ^g$ Department of Physics, University of Warwick, Coventry, United Kingdom\\
$ ^l$ Lawrence Berkeley National Laboratory, University of California, Berkeley, CA 94720, USA\\
$ ^s$ Institute of Particle and Nuclear Studies, High Energy Accelerator Research Organization (KEK), Tsukuba, 305-0801, Japan\\
$ ^t$ Universit\'e Paris-Saclay, CNRS/IN2P3, IJCLab, 91405 Orsay, France\\
}
\end{center}

\vspace{\fill}

\begin{abstract}
  \noindent
  The angle $\gamma$ of the Cabibbo--Kobayashi--Maskawa unitarity triangle is a benchmark parameter of the Standard Model of particle physics.
  A method to determine $\gamma$ from $\Bp \to D\Kp$ with subsequent $D \to \KS\pip\pim$ or similar multibody decays has been proven to provide good sensitivity.
  We review the first discussions on the use of this technique, and its impact subsequently.
  We propose that this approach should be referred to as the BPGGSZ method.  
\end{abstract}

\vspace{\fill}

\begin{center}
    The authors thank
    Alex~Bondar, Anton~Poluektov, Anjan~Giri, Yuval~Grossman, Abner~Soffer and Jure~Zupan
    for endorsing this note.
\end{center}

\vspace{\fill}

\end{titlepage}

\newpage
\setcounter{page}{2}
\mbox{~}

\cleardoublepage

\renewcommand{\thefootnote}{\arabic{footnote}}
\setcounter{footnote}{0}

\pagestyle{plain} 
\setcounter{page}{1}
\pagenumbering{arabic}

The angle $\gamma = {\rm arg}\left(-V_{ud} V^*_{ub}/V_{cd}V^*_{cb}\right)$ of the unitarity triangle formed from elements of the Cabibbo--Kobayashi--Maskawa quark mixing matrix~\cite{Cabibbo:1963yz,Kobayashi:1973fv} is a benchmark parameter of the Standard Model of particle physics.\footnote{
  An alternative notation, $\phi_3 \equiv \gamma$, is also in widespread use in the literature.}
The value of $\gamma$ is a measure of the extent to which the \CP symmetry between particles and antiparticles is violated in the weak interactions of quarks.
It can be determined with negligible theoretical uncertainty from measurements of decay rates and \CP-violating asymmetries in processes where interference between $b \to c\bar{u}s$ and $b \to u\bar{c}s$ transitions can occur~\cite{Bigi:1988ym}.
The archetypal example, which is also the most experimentally accessible, is that of $\Bp \to D\Kp$ decays, where $D$ indicates a neutral $D$ meson that is an admixture of $\Dz$ and $\Dzb$ states, but the same concepts are valid also for related channels such as $\Bp \to \Dstar\Kp$, $\Bp \to D\Kstarp$ and $\Bz \to D\Kstarz$.
The observables (relative decay rates and \CP-violating asymmetries) can be expressed in terms of $\gamma$, $r_B$ and $\delta_B$, where $r_B$ is the relative magnitude of the $b \to u\bar{c}s$ and $b \to c\bar{u}s$ transition amplitudes and $\delta_B$ is their relative strong (\ie\ \CP-conserving) phase.
Detailed reviews on the determination of $\gamma$ from such processes can be found, for example, in Refs.~\cite{PDG2020,HFLAV,Bevan:2014iga,LHCb-PAPER-2016-032,Gershon:2016fda}.

Initial discussions of the use of this approach to obtain experimental sensitivity to $\gamma$ focussed on the case where the neutral $D$ meson decays to a \CP\ eigenstate, such as (\CP-even) $\Kp\Km$, $\pip\pim$ or (\CP-odd) $\KS\piz$~\cite{Gronau:1990ra,Gronau:1991dp}.
In this approach, now widely known as the GLW method, the amplitudes for \Dz\ and \Dzb\ decays to the final states of interest are related trivially, under the assumption of negligible effects of \CP\ violation in the $D$ system.
The $B$ decay observables can then be expressed in terms of the unknown parameters $\left(\gamma,\ r_B,\ \delta_B\right)$ only.
To determine $\gamma$ from this method alone, however, requires sensitivity to observe the relatively small \CP\ asymmetries, expected to be comparable in magnitude to $r_B$, which is of order $0.1$ for $\Bp \to D\Kp$ decays.  
Moreover, such a determination of $\gamma$ suffers from trigonometric ambiguities since there are only three linearly independent observables in the GLW method, even in the case that both \CP-even and \CP-odd $D$ decay final states are used. 

To overcome these issues, it is essential to include $D$ decays to non-\CP eigenstates.
The use of $\Bp \to D\Kp$ with subsequent $D \to \Kmp\pipm$ decays was noted as being particularly valuable~\cite{Atwood:1996ci}, since both doubly Cabibbo-suppressed (\eg $\Dzb \to \Km\pip$) and Cabibbo-favoured (\eg\ $\Dz \to \Km\pip$) amplitudes contribute.
The smallness of the relative magnitude of these amplitudes, denoted $r_D$, complements the size of $r_B$, so that larger \CP\ asymmetries are possible in $\Bp \to D\Kp$ decays.
Both $r_D$ and the relative strong phase between the $D$ decay amplitudes, $\delta_D$, can be determined independently of the $B$ decay data, so that the observables depend on the same set of unknown parameters $\left(\gamma,\ r_B,\ \delta_B\right)$ as in the GLW case.
This approach is now widely known as the ADS method.
While the two-body $D \to \Kmp\pipm$ decays are particularly attractive experimentally, the ADS method can also be applied for multibody decays such as $D \to \Kmp\pipm\piz$ and $D\to\Kmp\pipm\pip\pim$~\cite{Atwood:1996ci,Atwood:2000ck}.

Before the start of the BaBar and Belle experiments, $\gamma$ was expected to be constrained only weakly by the anticipated data~\cite{Harrison:1998yr}.
As the first results with the GLW and ADS methods were published~\cite{Abe:2002ha,Swain:2003yu,Aubert:2003uy,Aubert:2004fb}, their precision confirmed that much larger data samples would be necessary to constrain $\gamma$ to the $10\degrees$ level or better.

This situation motivated the investigation of further decay modes that could be usefully employed to determine $\gamma$.
Several authors noted that the available statistics could be increased by using multibody final states, including those from doubly Cabibbo-suppressed~\cite{Atwood:2000ck} and singly Cabibbo-suppressed~\cite{Grossman:2002aq} $D$ decays.
However, these focussed either on inclusive approaches, in which the whole phase-space was integrated over, or on contributions from particular resonances.  
The key point of how interference between different resonances in the Dalitz plot of a multibody $D$ decay could be exploited to measure $\gamma$ was not realised until it was proposed to use decays into self-conjugate multibody final states, such as $\KS\pip\pim$.
This method has proven over time to have very good sensitivity.
The original work was performed separately and independently by two groups. 
Bondar and Poluektov developed their ideas within the Belle collaboration, while Giri, Grossman, Soffer and Zupan developed theirs for a theoretical paper. 

The first presentation on such a method was made by Alex Bondar at an internal Belle collaboration workshop in September 2002~\cite{Bondar}.
The slides of this presentation, which have not previously been available publicly, are included in Appendix~\ref{app:Bondar-talk}.
The idea was inspired by the Dalitz plot analysis of $\Dz \to \KS\pip\pim$ decays by the CLEO collaboration that had been published earlier in 2002~\cite{Muramatsu:2002jp}, in which it was demonstrated that the relative magnitude and phase of the doubly Cabibbo-suppressed $\Dz \to \Kstarp\pim$ and Cabibbo-favoured $\Dz \to \Kstarm\pip$ decay amplitudes could be determined from a sample of flavour-tagged $D$ mesons.
If the same relative phase could be measured separately in samples of neutral $D$ mesons from $\Bp \to D\Kp$ and charge-conjugate decays, the difference between these two quantities could be used to obtain $2\gamma$.
The solution thus obtained would be unique in the range $\left[ 0, \pi \right]$.
The concept was developed further, subsequently to the initial presentation, and implemented with the Belle data by Anton Poluektov.
Due to the strong competition between Belle and BaBar at that time, it was decided not to publish the method as a standalone paper, but rather to describe the approach together with the experimental results in a Belle collaboration publication.
The first results were presented in preliminary form at the Lepton Photon conference in August 2003~\cite{Abe:2003cn}, and published not long thereafter~\cite{Poluektov:2004mf}.

The first publication of the idea was the paper by Anjan Giri, Yuval Grossman, Abner Soffer, and Jure Zupan~\cite{Giri:2003ty}.
This was made available on the arXiv preprint server in March 2003, and presented at the CKM 2003 Workshop in early April, prior to publication.
In addition to discussing that $\gamma$ could be determined from a model-dependent fit to the Dalitz plot distribution of $D \to \KS\pip\pim$ decays produced in $\Bp \to D\Kp$ and charge conjugate processes,
Ref.~\cite{Giri:2003ty} proposed a model-independent approach based on binning the Dalitz plot.
This model-independent approach removes a potentially large source of uncertainty originating from the description of the {\it a priori} unknown strong phase variation across the Dalitz plot, at the cost of some statistical precision due to the finite size of the bins.
Each bin has associated with it a set of hadronic parameters, corresponding to the $r_D$ and $\delta_D$ parameters of the ADS method, that can in principle be determined independently of the $B$ decay data, so that again the only unknowns to be found are $\left(\gamma,\ r_B,\ \delta_B\right)$. 
In particular, Ref.~\cite{Giri:2003ty} introduced the $c_i$ and $s_i$ parameters, which are the amplitude-weighted averages of the cosine and sine of the strong phase difference between \Dz\ and \Dzb\ decay amplitudes within Dalitz plot bin $i$.
These parameters can be determined from quantum-correlated $\psi(3770) \to D\Dbar$ decays.

Both model-dependent and model-independent variants of the method to determine  $\gamma$ from $\Bpm \to D\Kpm$ with multibody $D$ decays, which we propose to refer to as the BPGGSZ method (see Appendix~\ref{app:cite} for a recommendation on appropriate citations), 
have been intensively pursued by experiments.\footnote{
    This approach has until now been referred to in a variety of ways, sometimes using the GGSZ acronym.
}
The most recent results from the BaBar~\cite{delAmoSanchez:2010rq}, Belle~\cite{Poluektov:2010wz} and LHCb~\cite{LHCb-PAPER-2014-017} collaborations with the model-dependent approach, using $D \to \KS\pip\pim$ decay models obtained in Refs.~\cite{delAmoSanchez:2010xz,Poluektov:2006ia},\footnote{
  An updated study of the $D \to \KS\pip\pim$ decay amplitude has been published in Ref.~\cite{Adachi:2018jqe}, but not yet used in any $\gamma$ determination. 
} 
have statistical uncertainties on $\gamma$ as low as $12$--$15\degrees$ and model uncertainties that vary in the range $3$--$9\degrees$ depending on how conservative a range of model variations is considered.
The latest results from the Belle~\cite{Aihara:2012aw} and LHCb~\cite{LHCb-PAPER-2014-041,LHCb-PAPER-2018-017} collaborations with the model-independent approach have uncertainties on $\gamma$ of $10$--$15\degrees$.
A small contribution to this, around $4\degrees$, is due to the limited precision with which the $c_i$ and $s_i$ parameters have been measured using data from the CLEOc experiment~\cite{Briere:2009aa,Libby:2010nu} following the scheme for binning of the Dalitz plot proposed in Refs.~\cite{Bondar:2005ki,Bondar:2008hh}.\footnote{
  More precise measurements of the $c_i$ and $s_i$ parameters have recently been reported by the BESIII collaboration~\cite{Ablikim:2020yif,Ablikim:2020lpk}, but not yet used in any $\gamma$ determination.
}
Further improvement in precision can be anticipated as the existing LHCb data sample is analysed, and as much larger data samples are collected in the future with upgrades of the LHCb detector~\cite{LHCb-TDR-012,LHCb-PAPER-2012-031,CERN-LHCC-2017-003,LHCb-PII-Physics} and with the Belle~II experiment~\cite{Abe:2010gxa,Kou:2018nap}.

In addition to its use in the $\Bp \to D\Kp$, $D \to \KS\pip\pim$ decay chain, the BPGGSZ method has also been used with $\Bp \to D^*\Kp$~\cite{delAmoSanchez:2010rq,Poluektov:2010wz}, $\Bp \to D\Kstarp$~\cite{delAmoSanchez:2010rq,Poluektov:2006ia} and $\Bz \to D\Kstarz$~\cite{Negishi:2015vqa,LHCb-PAPER-2016-006,LHCb-PAPER-2016-007} decays. 
In addition, $D$ decays to the $\KS\Kp\Km$~\cite{delAmoSanchez:2010rq,LHCb-PAPER-2014-041,LHCb-PAPER-2018-017} and $\KS\pip\pim\piz$~\cite{K:2017qxf,Resmi:2019ajp} final states have been exploited and application of the method with other self-conjugate multibody decays is likely in the future.
Moreover, the BPGGSZ method has inspired similar methods to determine additional important quantities in heavy flavour physics.
Analysis of $B \to D\piz$ and similar decays with $D \to \KS\pip\pim$ and other self-conjugate multibody final states can be used to determine the angle $\beta$ of the unitarity triangle~\cite{Bondar:2005gk}; this method has been implemented in both model-dependent~\cite{Adachi:2018itz,Adachi:2018jqe} and model-independent~\cite{Vorobyev:2016npn} variants.
A ``double Dalitz plot'' analysis for $B \to D\pip\pim$ with $D \to \KS\pip\pim$ decays has also been proposed~\cite{Bondar:2018gpb}, building on a similar concept for $\gamma$ determination~\cite{Gershon:2009qr,Craik:2017dpc}.
Binning of the $D \to \KS\pip\pim$ Dalitz plot has also been noted to have highly advantageous feature for the experimental determination of \D\ meson mixing parameters from these decays~\cite{Bondar:2010qs,Thomas:2012qf,DiCanto:2018tsd}. 
Results obtained with these methods~\cite{LHCb-PAPER-2015-042,LHCb-PAPER-2019-001} currently provide the best sensitivity out of all charm mixing measurements to the mass difference between the neutral \D\ meson eigenstates.

In conclusion, the development of the BPGGSZ method 
enabled significant improvement in the measurement of $\gamma$, removing ambiguities in its determination.
This permitted global fits to the parameters of the CKM matrix, allowing for, and deriving stringent bounds on, contributions from physics beyond the SM in loop-dominated processes~\cite{Charles:2004jd,Bona:2005eu}.
Such analyses proved that the dominant contributions to flavour-changing processes and to \CP\ violation in meson decays are those of the Standard Model, as recognised by the award of the 2008 Nobel Prize in Physics to Kobayashi and Maskawa.  
Results obtained with the BPGGSZ method continue to provide some of the most precise constraints on $\gamma$ that enter today's global fits~\cite{HFLAV,PDG2020}, and are expected to continue to do so in the future.

\section*{Acknowledgements}

The authors gratefully acknowledge Alex Bondar, Anton Poluektov, Anjan Giri, Yuval Grossman, Abner Soffer and Jure Zupan for providing background information, and for comments on the manuscript.
The authors would also like to thank S\'{e}bastien Descotes-Genon, Jim Libby and Sneha Malde for helpful discussions.

\addcontentsline{toc}{section}{References}
\setboolean{inbibliography}{true}
\bibliographystyle{LHCb}
\bibliography{references,main}

\clearpage
\appendix
\section{First presentation of method to determine $\gamma$ from $\Bp \to D\Kp$, $D\to \KS\pip\pim$ decays}
\label{app:Bondar-talk}

\begin{figure}[!hb]
  \centering
  \includegraphics[width=0.85\textwidth]{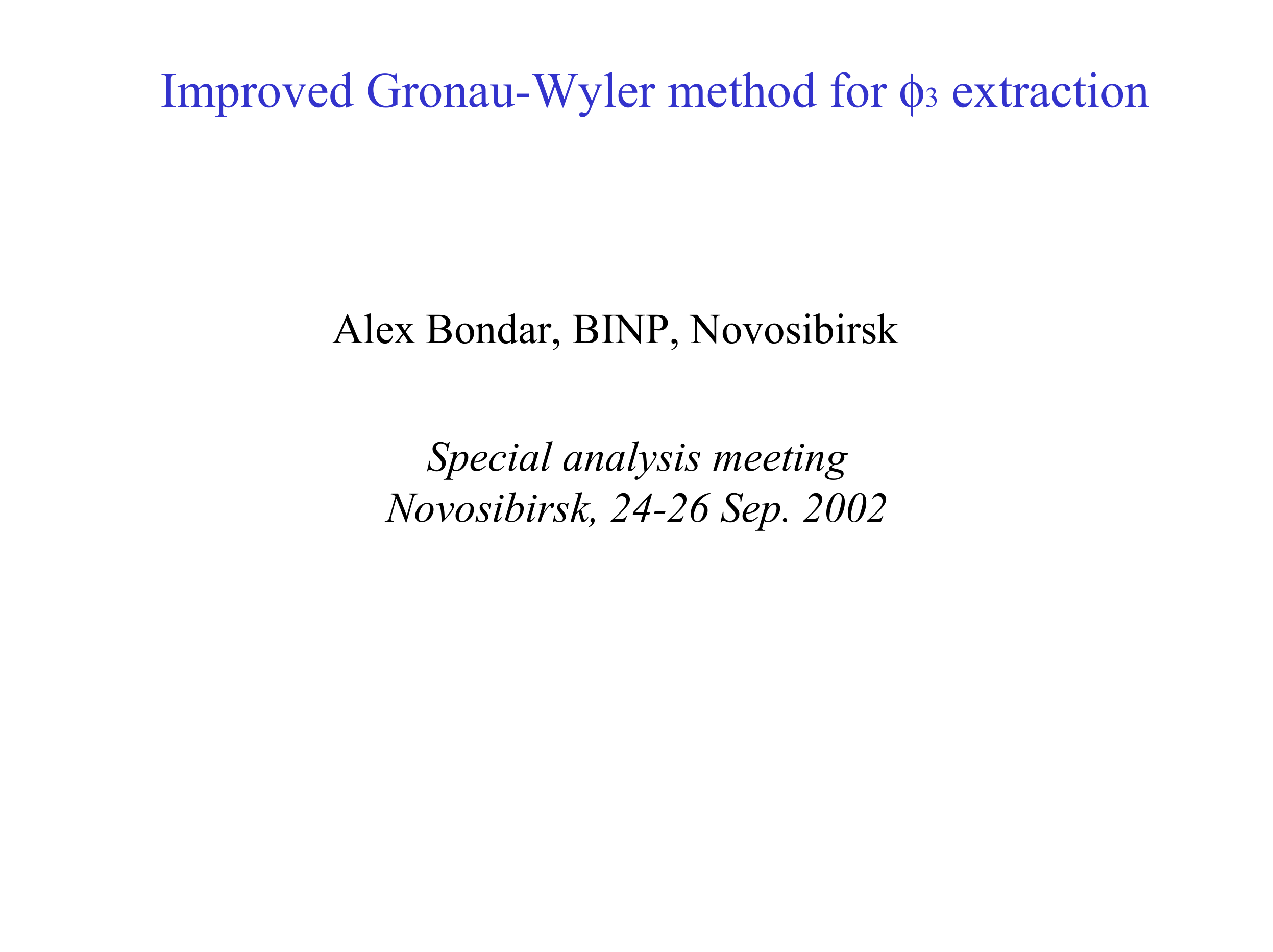} \\
  \vspace{-3.7cm}
  \rule{\linewidth}{0.5pt}\\
  \includegraphics[width=0.85\textwidth]{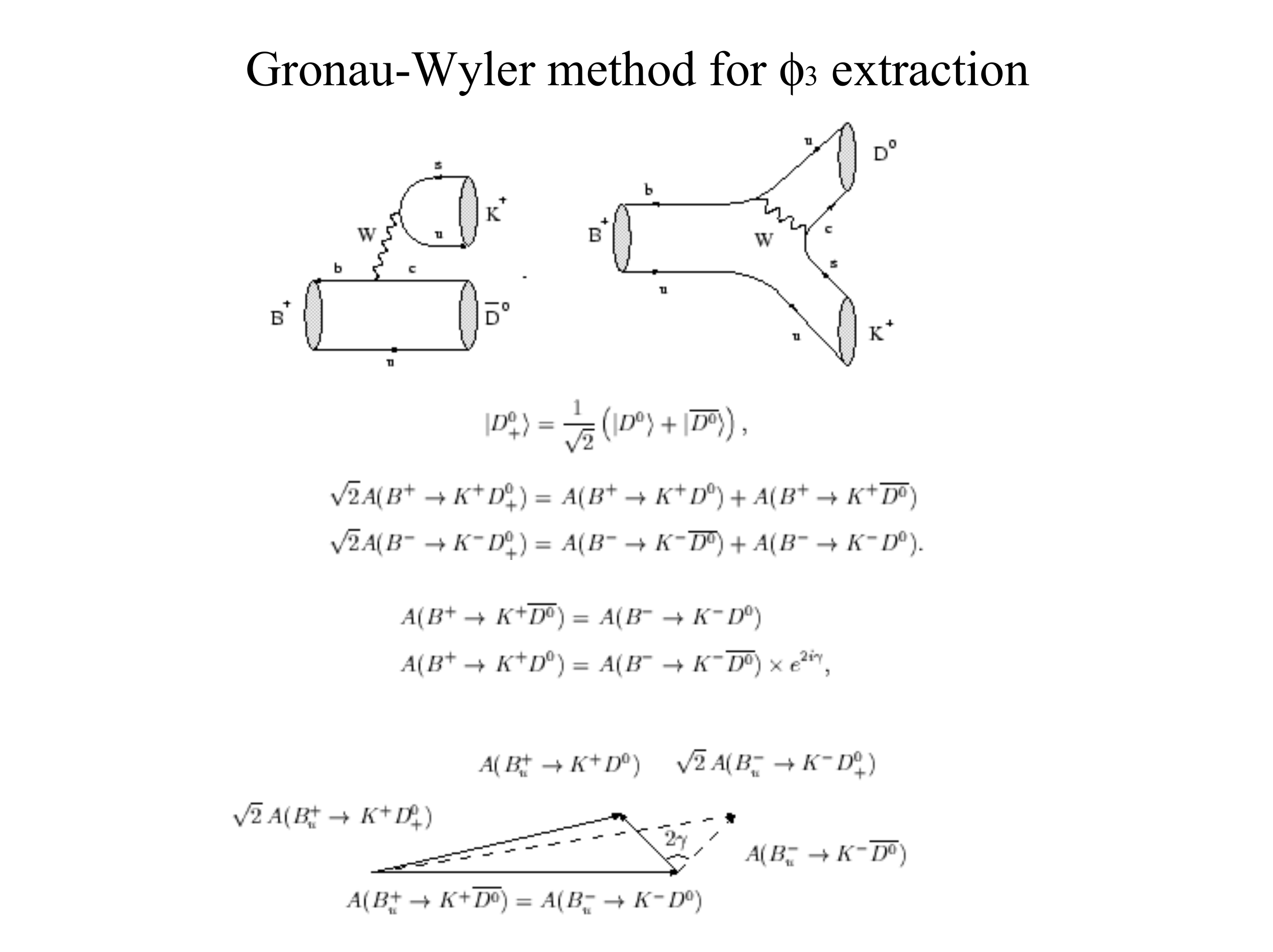}
  \caption{
    Slides 1--2 of the presentation of Ref.~\cite{Bondar}.
    \label{fig:bondar1}
  }
\end{figure}

\begin{figure}[!htb]
  \centering
  \includegraphics[width=0.85\textwidth]{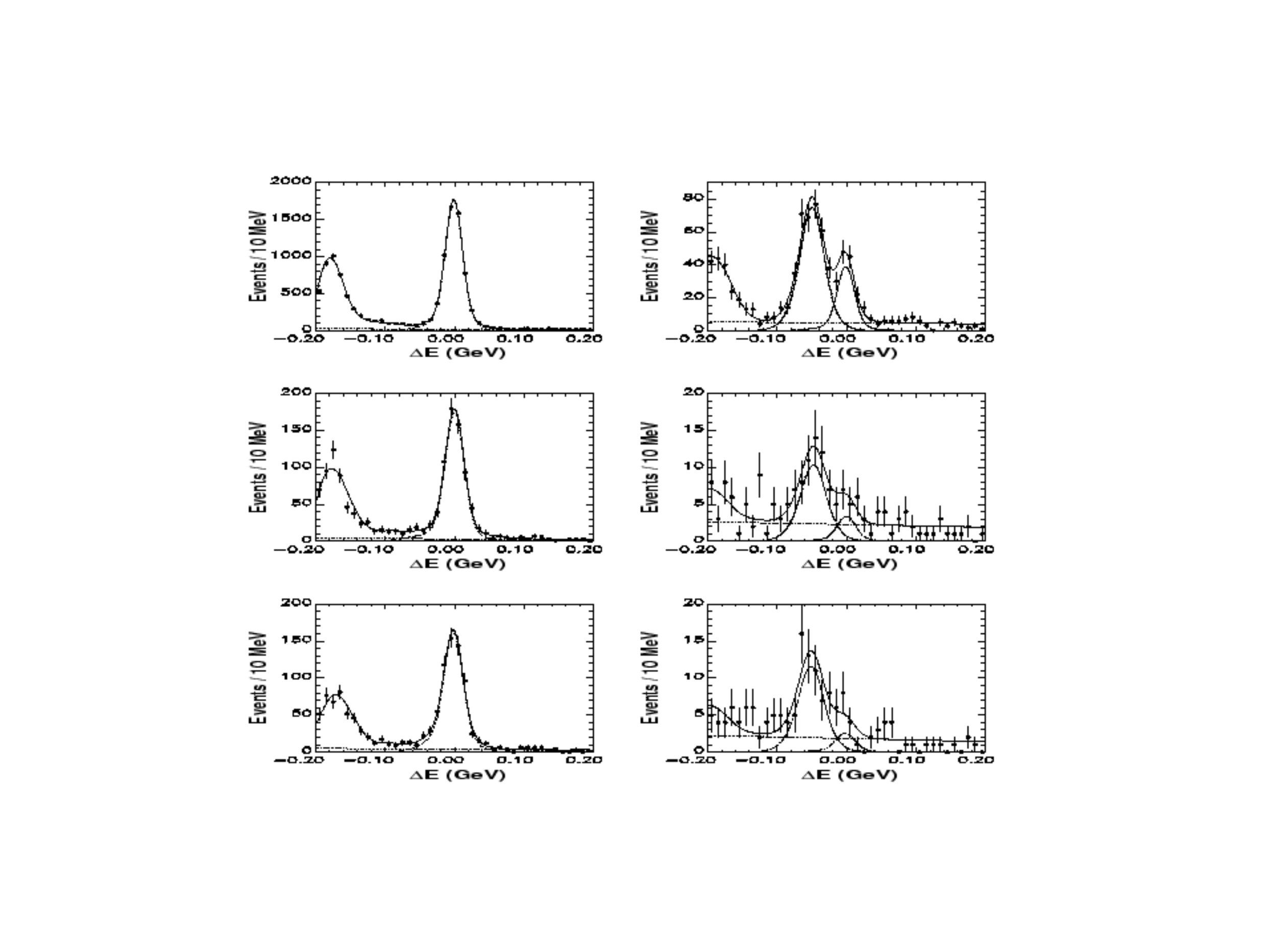}\\
  \rule{\linewidth}{0.5pt}\\
  \includegraphics[width=0.85\textwidth]{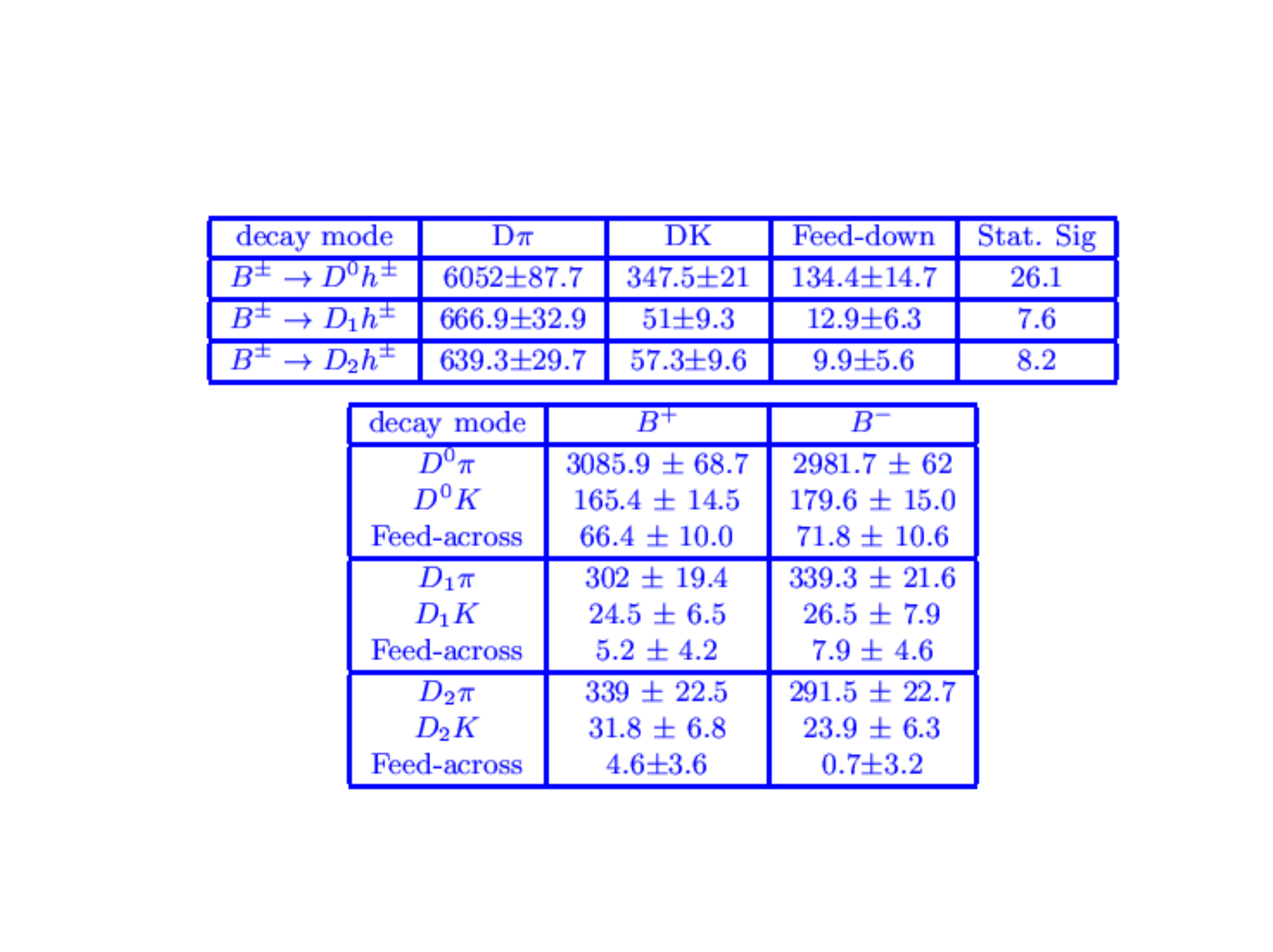}
  \caption{
    Slides 3--4 of the presentation of Ref.~\cite{Bondar}.
    \label{fig:bondar2}
  }
\end{figure}

\begin{figure}[!htb]
  \centering
  \includegraphics[width=0.85\textwidth]{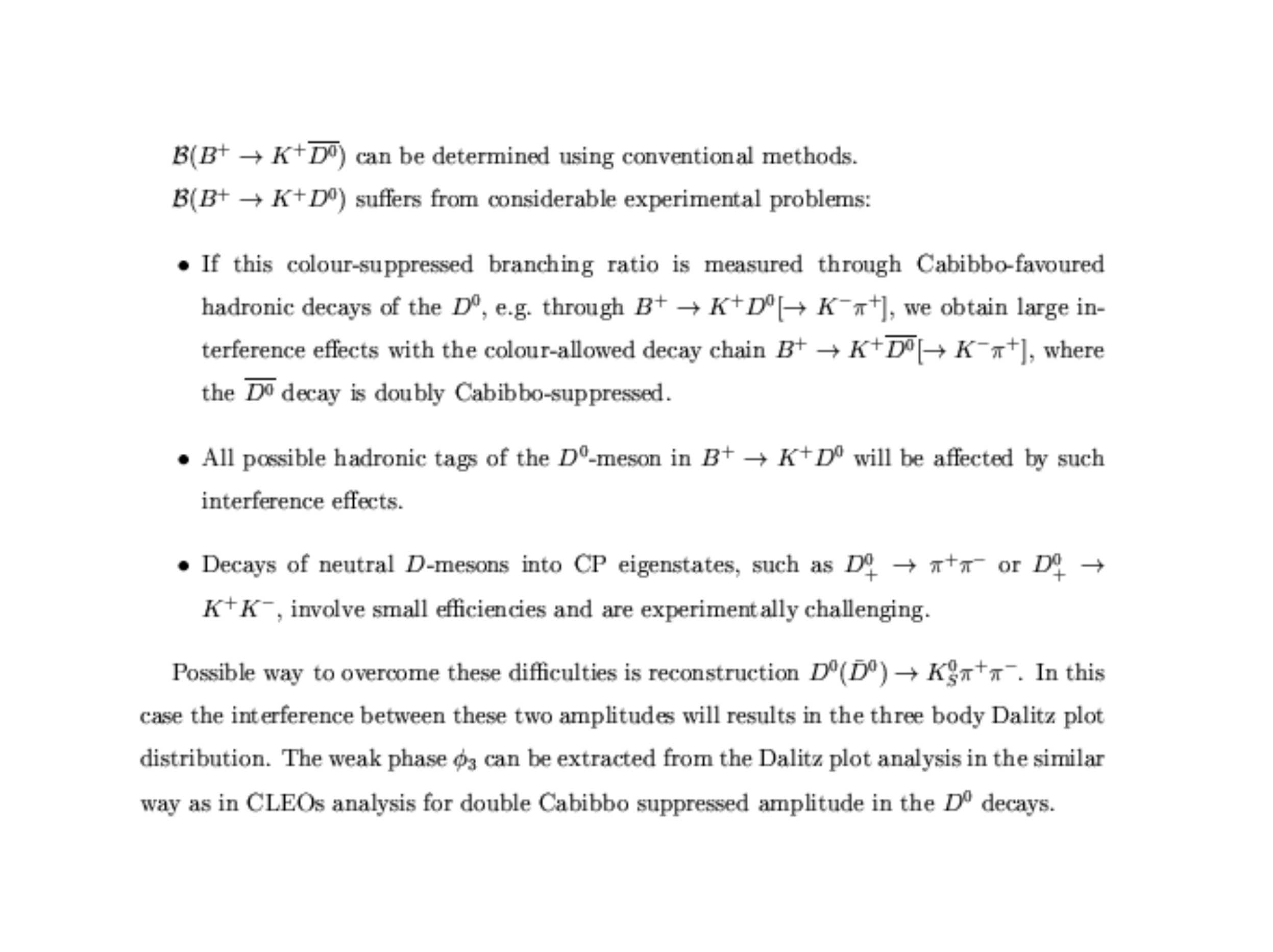}\\
  \rule{\linewidth}{0.5pt}\\
  \includegraphics[width=0.85\textwidth]{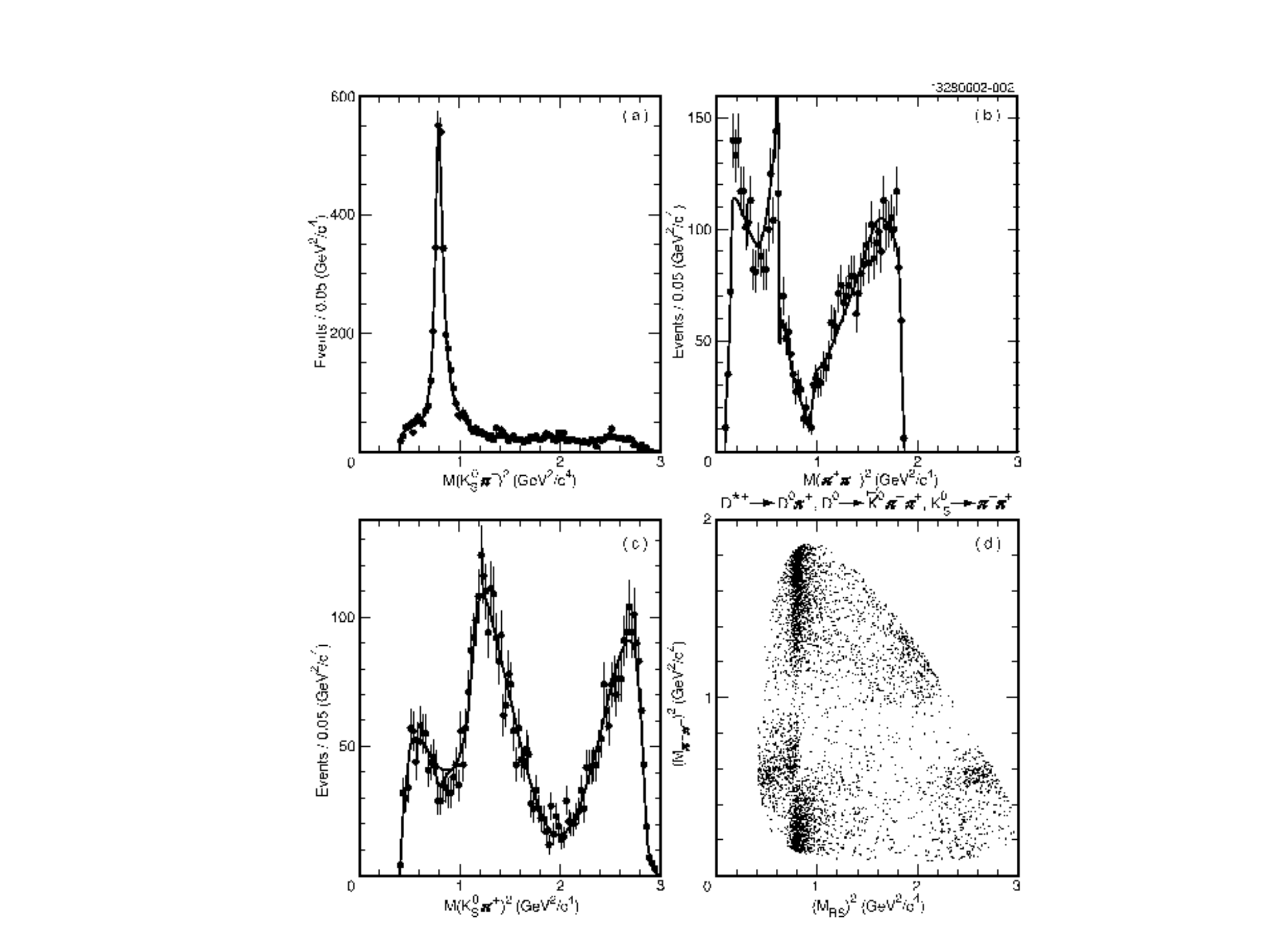}
  \caption{
    Slides 5--6 of the presentation of Ref.~\cite{Bondar}.
    \label{fig:bondar3}
  }
\end{figure}

\begin{figure}[!htb]
  \centering
  \includegraphics[width=0.85\textwidth]{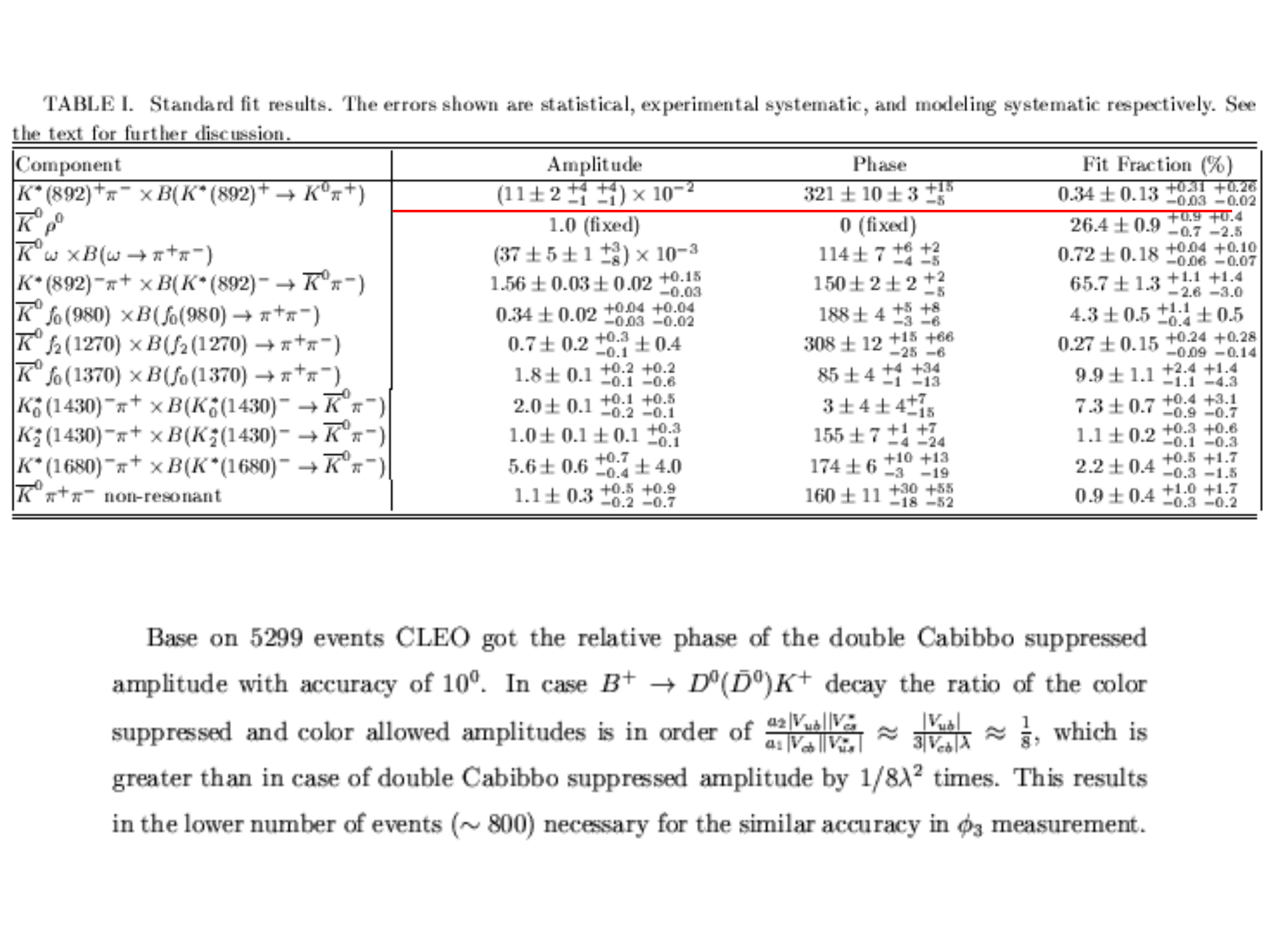}\\
  \rule{\linewidth}{0.5pt}\\
  \includegraphics[width=0.85\textwidth]{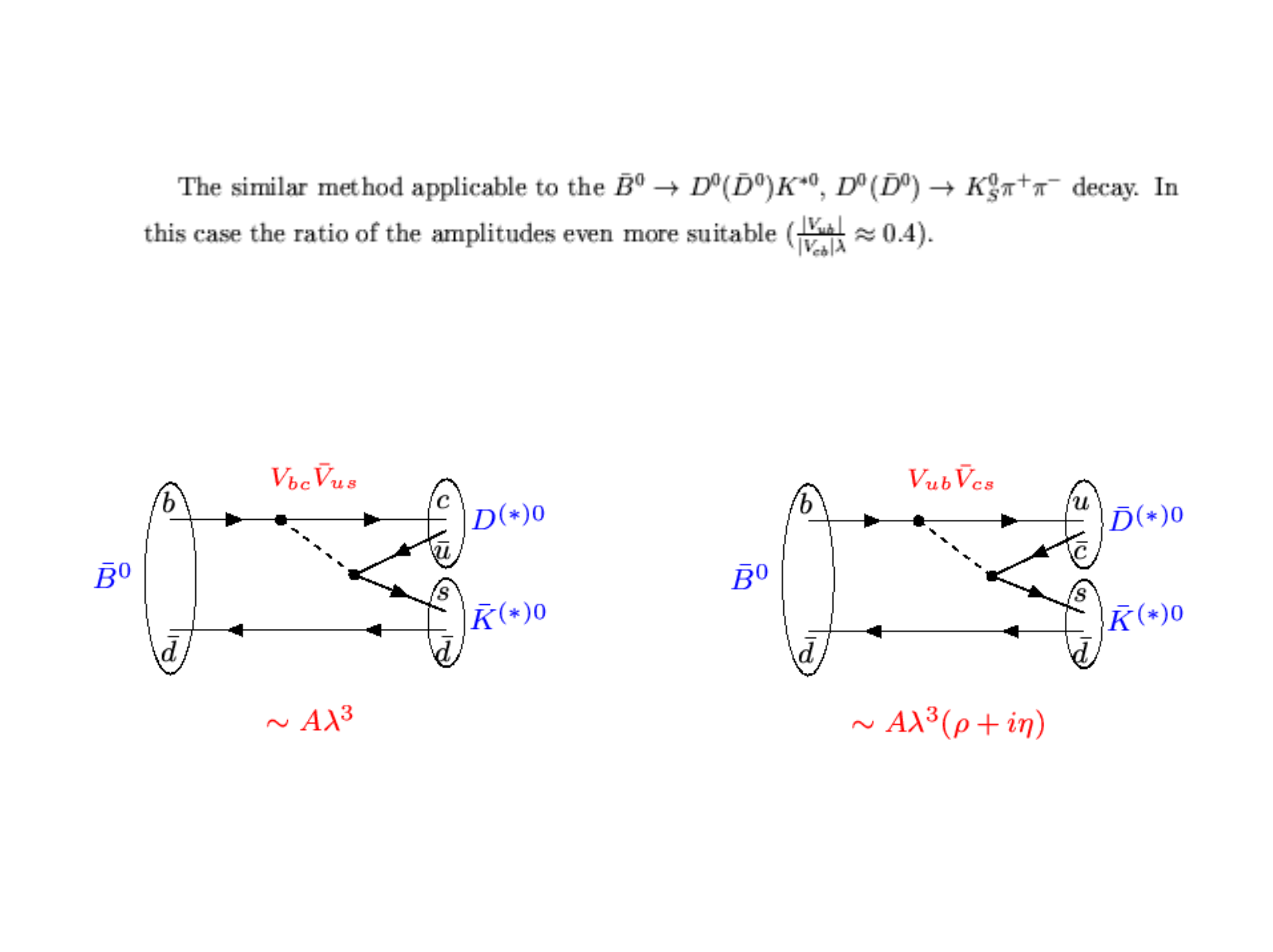}
  \caption{
    Slides 7--8 of the presentation of Ref.~\cite{Bondar}.
    \label{fig:bondar4}
  }
\end{figure}

\begin{figure}[!htb]
  \centering
  \includegraphics[width=0.85\textwidth]{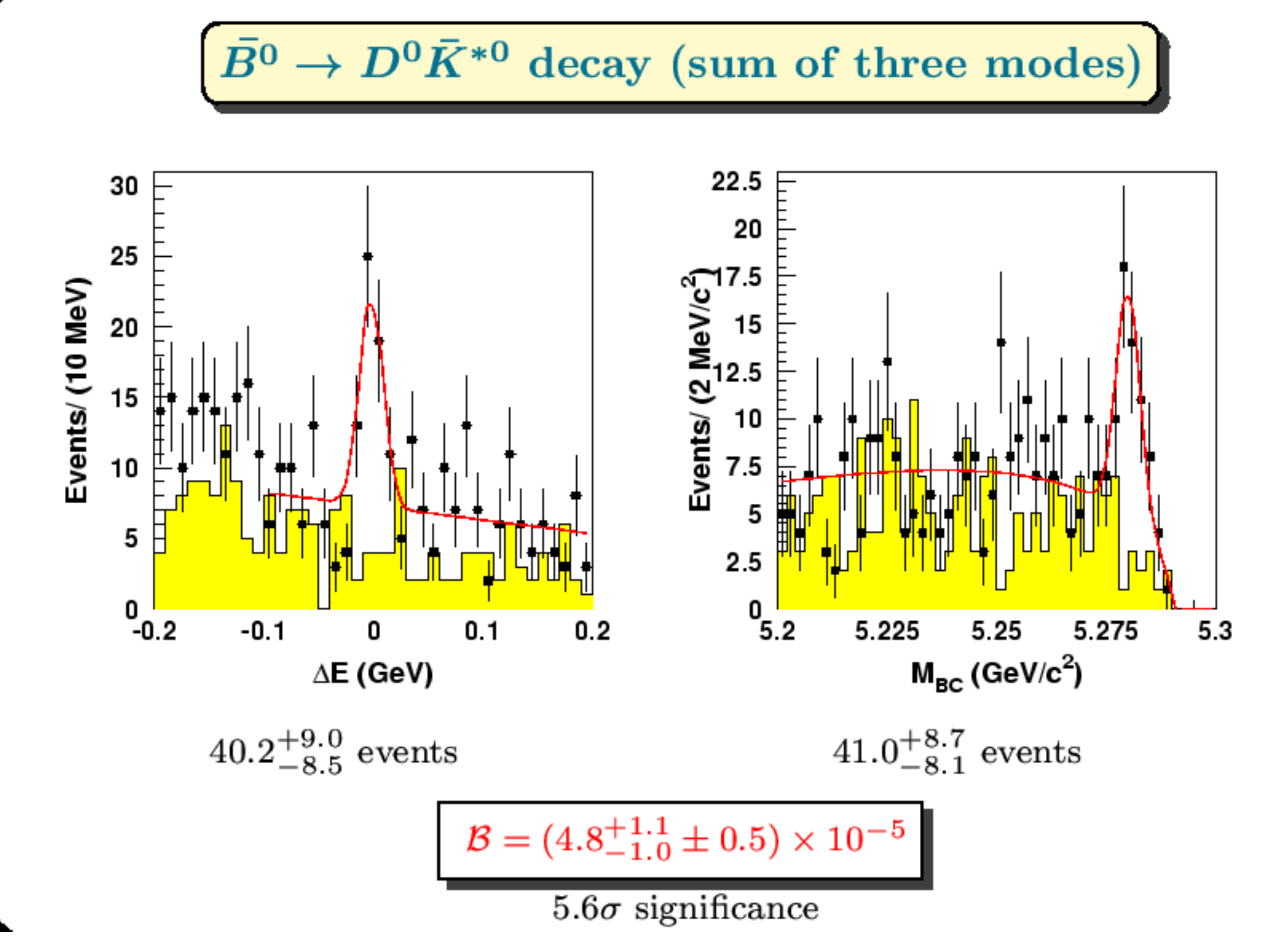}\\
  \rule{\linewidth}{0.5pt}\\
  \includegraphics[width=0.85\textwidth]{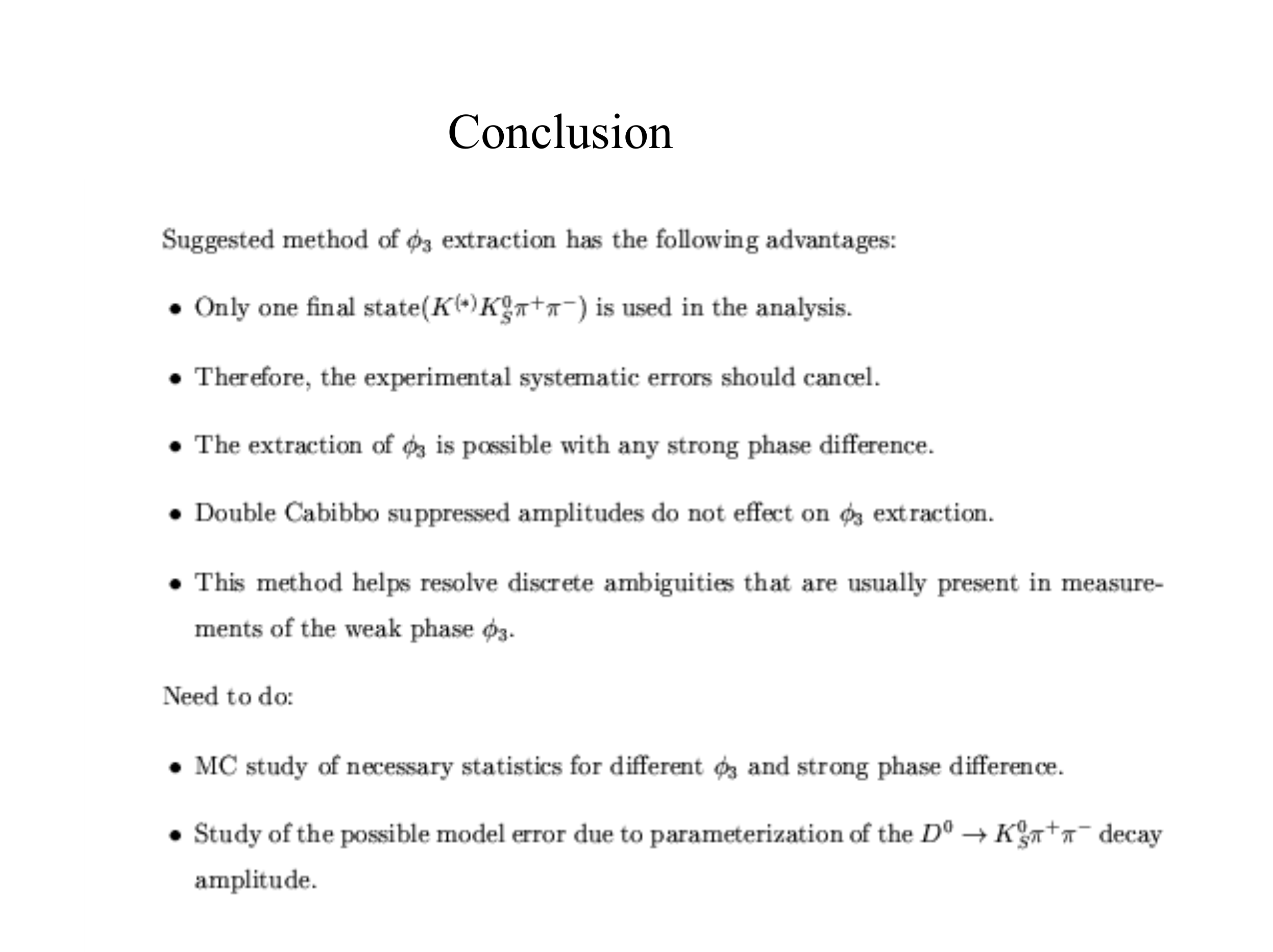}
  \caption{
    Slides 9--10 of the presentation of Ref.~\cite{Bondar}.
    \label{fig:bondar5}
  }
\end{figure}

Figures~\ref{fig:bondar1}--~\ref{fig:bondar5} contain the slides of the presentation at the Belle collaboration workshop in September 2002~\cite{Bondar}, in which the concept of measuring $\gamma$ (denoted $\phi_s$ in the slides) from $\Bp \to D\Kp$, $D\to \KS\pip\pim$ decays was first set out.

\clearpage
\section{Recommendation for citation of BPGGSZ method}
\label{app:cite}

We recommend to use the following citations for the BPGGSZ method:

\begin{center}
\begin{tabular}{c p{14.5cm}}
[1] & A.~Bondar. Proceedings of BINP special analysis meeting on Dalitz analysis, 24--26 Sep.~2002, unpublished. \\ \relax
[2] & A.~Giri, Y.~Grossman, A.~Soffer, and J.~Zupan, Determining $\gamma$ using $\Bp \to D\Kpm$ with multibody $D$ decays,
\href{https://doi.org/10.1103/PhysRevD.68.054018}{Phys.\ Rev.\ \textbf{D68} (2003) 054018},
\href{http://arxiv.org/abs/hep-ph/0303187}{{\normalfont\ttfamily arXiv:hep-ph/0303187}}. \\ \relax
[3] & Belle collaboration, A.~Poluektov et al., Measurement of $\phi_3$ with Dalitz plot analysis of $\Bpm \to D^{(*)}\Kpm$ decay,
\href{https://doi.org/10.1103/PhysRevD.70.072003}{Phys.\ Rev.\ \textbf{D70} (2004) 072003},
\href{http://arxiv.org/abs/hep-ex/0406067}{{\normalfont\ttfamily arXiv:hep-ex/0406067}}.
\end{tabular}
\end{center}

\end{document}